\documentclass[aps,prl,twocolumn,showpacs,superscriptaddress,amssymb]{revtex4}
\usepackage{graphicx}
\usepackage{hyperref}
\usepackage{natbib}
\usepackage{bm}
\begin{document}
\title{Numerical modeling of the central spin problem using the 
spin coherent states P-representation}
\author{K. A. Al-Hassanieh}
\affiliation{Condensed Matter Sciences Division, Oak Ridge National Laboratory, Oak Ridge, TN 37996}
\affiliation{Department of Physics, University of Tennessee, Knoxville TN 37831, USA}
\affiliation{NHMFL and Department of Physics, Florida State University, Tallahassee, FL 32306, USA}
\author{V. V. Dobrovitski}
\affiliation{Ames Laboratory and Department of Physics and Astronomy, 
  Iowa State University, Ames IA 50011, USA}
\author{E. Dagotto}
\affiliation{Condensed Matter Sciences Division, Oak Ridge National Laboratory, Oak Ridge, TN 37996}
\affiliation{Department of Physics, University of Tennessee, Knoxville TN 37831, USA}
\author{B. N. Harmon}
\affiliation{Ames Laboratory and Department of Physics and Astronomy, 
  Iowa State University, Ames IA 50011, USA}
\begin{abstract}
In this work, we consider decoherence of a central spin by a
spin bath. 
%Theoretical understanding of this process is timely
%and relevant for various experiments, such as the recent study of
%decoherence of the electron spin in a quantum dot by the nuclear
%spins (A.~C.~Johnson et al., Nature {\bf 435}, 925 (2005)). 
In order to study the non-perturbative decoherence regimes, we
develop an efficient mean-field-based method for
modeling the spin-bath decoherence, based on the P-representation
of the central spin density matrix. The method can be applied to
longitudinal and transversal relaxation at different external
fields. In particular, by modeling large-size quantum systems
(up to 16000 bath spins), we make controlled predictions for the
slow long-time decoherence of the central spin.
\end{abstract}

\maketitle

%Any real physical system interacts with the outside world.
%Such an interaction leads to decoherence: the
%initial quantum state of the system quickly decays into an incoherent
%mixture of several states \cite{vonNeumann}. 
Detailed
understanding of the decoherence of various quantum systems is important
for many areas, from quantum optics and solid-state physics to
the quantum information processing, where decoherence constitutes
a major obstacle for building a practical quantum computer.
For example, in a quantum dot-based architecture, the quantum bit is
represented by a spin of a single electron (central spin) 
placed in a quantum dot (QD). Due to interaction with the bath of 
nuclear spins in a QD, the electron spin quickly ``looses memory''
about its initial orientation and can not be used for computation.
%Similar processes plague other spin-based architectures (shallow donors in Si
%\cite{kane}, nuclear spins in liquid- and solid-state NMR \cite{chuang},
%etc.). 
Experimental studies of this process has become possible very
recently \cite{johnson,taylor}, and detailed theoretical
understanding of the experimental results is timely and
important. Moreover, the problem of a central spin coupled to
the spin bath \cite{gaudin} (the ``quantum central spin
problem'') has recently arised in other contexts (decoherence in 
magnetic molecules, dynamics of the cold fermions
pairing), and has attracted much attention.

Decoherence is a complex quantum many-body phenomenon,
and satisfactory solutions can be obtained only in
very special cases \cite{khaetskii,efros}. 
Perturbation theory can be successfully applied in the case of a strong magnetic 
field or polarized nuclear spin bath (which produces a strong Overhauser field acting
on the central spin) \cite{loss,hu}.
But for the experimentally important
non-perturbative regimes, no well-justified method, numerical or
analytical, has been suggested yet.
The static approximation \cite{dkdh,efros,akakii},
which simply neglects the dynamics of the bath spins, works only
at short times, while the interesting long-time dynamics of the central spin
remains an open problem, and the simulations on moderate-sized systems
($\sim 20$ spins) do not give conclusive results.

In this work, we present a novel approach to the quantum central 
spin problem based on the time-dependent mean field (TDMF) theory. 
It has been pointed out previously \cite{nazarov,emil}
that the mean field approach should be adequate, since the central spin
interacts with a large number $N$ of the bath spins (loosely speaking, 
the number of the ``nearest neighbors'' for the central 
spin is large, i.e.\ the coupling of the central spin with the bath has 
an infinite range). However, the standard mean field approach \cite{dirac} 
gives a clearly wrong answer (see below). We use the spin coherent state 
P-representation \cite{gardiner}
to modify the standard TDMF, and present an efficient approach, 
which gives excellent agreement 
with the exact solution of the many-spin quantum problem already for
rather small systems (up to 20 spins). By applying this method to
large-scale problems, we study the interesting long-time dynamics
of the central spin. Moreover, the P-representation approach
allows us to understand why the properly corrected TDMF theory works
for large $N$, and what the limitation of the method are. 

It is interesting to point out that the spin coherent states are
traditionally used in the spin path integrals and in the
semiclassical approximation for quantum spins. The powerful
methods based on P- and Q-representations, so useful in quantum
optics, have not been widely applied to the spin systems studies.
We expect that this work will help developing novel coherent-state 
approaches to studying spin systems. Another interesting point
concerns the basic ideas of the mean field theory. The standard 
mean field theory approximates, in the optimal way, the exact 
many-spin wavefunction as a product of single-spin wavefunctions.
However, when studying decoherence, we are only interested in the
state of the central spin, not the whole many-body state.
It is interesting that there is a justified
modification of the standard TDMF (presented below) which provides
a much better approximation for the relevant observables (the state of 
the central spin) at the expense of irrelevant information
(the state of the bath).

The electron spin in a QD interacting with
the bath of $N$ nuclear spins can be described by the Hamiltonian
\begin{equation}
{\cal H} = g_e^*\mu_B B_0 S_0^z + 
  \sum_{k=1}^N A_k {\mathbf S}_k{\mathbf S}_0=H_0 S_0^z + 
  \sum_{k=1}^N {\cal H}_k
\label{hamt}
\end{equation}
where ${\mathbf S}_0=(S_0^x,S_0^y,S_0^z)$ are the operators of the electron spin, 
${\mathbf S_k}$ are the operators of the bath spins, 
and $A_k=(8\pi/3)g_e^*\mu_Bg_n\mu_n |\Psi({\mathbf x}_k)|^2$ 
is the contact hyperfine coupling which is determined by the electron density
$|\Psi({\mathbf x}_k)|^2$ at the site ${\mathbf x}_k$ of the $k$-th nuclear
spin and by the Land\'e factors of the electron $g_e^*$ and of the nuclei $g_n$. 
The first term in Eq.~\ref{hamt} describes the Zeeman energy
of the electron spin in the external magnetic field $B_0$, and
the second term represents the contact hyperfine coupling.
The omitted terms, such as the Zeeman energy
of the nuclear spins and the non-isotropic part of the hyperfine coupling,
are very small and can be neglected for a wide range of experimental
situations. Eq.~\ref{hamt} is the standard Hamiltonian of the quantum
central spin problem \cite{gaudin}.

We are interested in the dynamics of the central spin, i.e.\ in the
dynamics of ${\mathbf s}_0(t) = {\mathrm{Tr}}{\rho(t){\mathbf S}_0}$, where
$\rho(t)$ is the density matrix of the total system (central spin plus the bath).
Although the quantum central spin problem is integrable, the formal solution
\cite{gaudin} is very complex, and, to our knowledge, it has not been used 
for actual calculations of ${\mathbf s}_0(t)$,
neither analytically, nor numerically. Efficient approximate approaches
are needed, and the TDMF theory is a natural first step.
Within the mean-field approach, we approximate the wavefunction $|\Psi\rangle$
of the total system as a product 
$|\Psi\rangle=|\psi_0\rangle\bigotimes_{k=1}^N |\psi_k\rangle$
of the single-spin wavefunctions $|\psi_j\rangle$.
The TDMF equations of motion for $|\psi_j\rangle$ are obtained by substituting
this ansatz into the Dirac's functional 
$D=\int dt \langle\Psi|i(d\Psi/dt) - {\cal H}\Psi\rangle$,
and requiring that $\delta D=0$ with respect to small variations of all $|\psi_k\rangle$.
The resulting equations of motion can be presented in a simple form as 
\begin{eqnarray}
\dot {\mathbf s}_j &=& [{\mathbf h}_j\times {\mathbf s}_j]\qquad (j=0\dots N)
\label{eoms}\\
{\mathbf h}_0&=&H_0{\mathbf e}_z+\sum_k A_k {\mathbf s_k},\;\;
{\mathbf h}_k= A_k {\mathbf s}_0\;\; (k=1\dots N)
\label{heff}
\end{eqnarray}
where ${\mathbf s}_j={\mathrm{Tr}}{\rho(t){\mathbf S}_j}$, and
$[{\mathbf h}_j\times {\mathbf s}_j]$ is the vector product of
${\mathbf h}_j$ and ${\mathbf s}_j$.
The physical meaning of these equations is simple: every $j$-th spin precesses in its 
own time-dependent effective field ${\mathbf h}_j$ given by 
Eq.~\ref{heff}.
However, TDMF theory gives a very bad approximation to the 
exact solution of the problem. It can be seen, for example, 
%by
%considering the case of equal couplings $A_k=A$, which is easy
%to solve analytically. The same conclusion can be reached 
from comparison of the standard TDMF theory with the exact numerical
solutions \cite{ddr} for several distributions of $A_k$ (see Fig.~\ref{fig2}):
the disagreement is significant.

In order to construct a working approximation based on the TDMF, let us 
consider the P-representation of the system's density matrix in the basis
of spin coherent states \cite{klauder,gardiner}. The spin coherent
state for spin $J$ is defined as 
$|\mu\rangle={\cal N}\sum_{m=0}^{2J} [(2J)!/(m!(J-m)!)]^{1/2}\mu^m |J-m\rangle$,
where ${\cal N}=(1+|\mu|^2)^{-J}$ is the normalization constant.
For a spin 1/2, the coherent state has a simple form 
$|\mu\rangle=\cos{(\theta/2)}|\uparrow\rangle +
\sin{(\theta/2)}{\mathrm e}^{i\phi}|\downarrow\rangle$,
where we used the parametrization 
$\mu=\tan{(\theta/2)}{\mathrm e}^{i\phi}$.
The basis of coherent states is overcomplete, and 
by chosing an appropriate real-valued function $\tilde A(\theta,\phi)$,
any hermitian operator $A$ can be represented in a diagonal form:
$A=\int_{\Omega} \tilde A(\theta,\phi) |\mu\rangle\langle\mu| 
 \sin{\theta}d\theta d\phi$, where the integration is performed over
the sphere. Note that 
$\tilde A(\theta,\phi)\neq \langle\theta,\phi|A|\theta,\phi\rangle$.
Moreover, the function $\tilde A(\theta,\phi)$ is not uniquely determined;
if we add to it any linear combination of the spherical harmonics $Y_l^m(\theta,\phi)$
of the order $l>2J$, then the value of the integral remains the same,
and the new function would define the same operator $A$.
Obviously, the many-spin density matrix $\rho$ can also be written in the 
diagonal form:
\begin{equation}
\rho(t) = \int p(\{\theta_j,\phi_j\},t) \bigotimes_{j=0}^N
  |\mu_j\rangle\langle\mu_j| 
  \prod_{j=0}^N \sin{\theta_j} d\theta_j d\phi_j,\
\label{prep}
\end{equation}
with a real-valued function $p(\{\theta_j,\phi_j\},t)$,
where $\{\theta_j,\phi_j\}$ is the set of all $\theta_0,\dots\theta_N$ and
$\phi_0,\dots\phi_N$. This representation for the density matrix is called
the P-representation \cite{gardiner}. Note that the operator
part of the expression (\ref{prep}) has a mean-field form, i.e.\ it is
a product of single-spin density matrices $|\mu_j\rangle\langle\mu_j|$.
In P-representation, the quantum-mechanical average $x={\mathrm{Tr}}(\rho(t)X)$
of any observable
$X$ can be calculated by a simple formula
\begin{equation}
x = \int p(\{\theta_j,\phi_j\},t) \bigotimes_{j=0}^N
  \langle\mu_j|X|\mu_j\rangle
  \prod_{j=0}^N \sin{\theta_j} d\theta_j d\phi_j.\
\label{observ}
\end{equation}

Our goal is to model the evolution of the function
$p(\{\theta_j,\phi_j\},t)$. However, the direct solution of the 
complex partial differential equation
for $p(\{\theta_j,\phi_j\},t)$ is impossible for a large (hundreds or more) 
number of spins, and we use a different approach. 
We note that if $p(\{\theta_j,\phi_j\},t)\ge 0$ then
this function can be interpreted as a probability for the system to be in
the product state $|\Psi\rangle=\bigotimes_{j=0}^N |\mu_j\rangle$, and
we need to simulate the dynamics of the probability distribution
$p(\{\theta_j,\phi_j\},t)$. To do that, we initially generate 
many realizations of the random
vector $(\theta^{(m)}_0,\dots\theta^{(m)}_N,\phi^{(m)}_0,\dots\phi^{(m)}_N)$ 
distributed according to 
the probability distribution $p(\{\theta_j,\phi_j\},0)$
(the index $m=1,\dots M$ enumerates the different realizations).
Then we propagate every initial vector 
$(\theta^{(m)}_0,\dots\theta^{(m)}_N,\phi^{(m)}_0,\dots\phi^{(m)}_N)$
in time, so that after a lapse of time $t$, the initial vector
evolves into a vector 
$(\Theta^{(m)}_0(t),\dots\Theta^{(m)}_N(t),\Phi^{(m)}_0(t),\dots\Phi^{(m)}_N(t))$.
If the equations of motion for all the variables $\Theta^{(m)}_j(t)$ and
$\Phi^{(m)}_j(t)$ are chosen correctly, then the function
$p(\{\theta_j,\phi_j\},t)=p({\Theta_j(t),\Phi_j(t)})$, and the value 
$x={\mathrm{Tr}}(\rho(t)X)$ of any observable $X$ can be calculated as
an average over all realizations:
%\begin{equation}
$x = \frac{1}{M}\sum_{m=1}^M \sin{\Theta^{(m)}_j}\bigotimes_{j=0}^N
  \langle\Theta^{(m)}_j,\Phi^{(m)}_j|X|\Theta^{(m)}_j,\Phi^{(m)}_j\rangle$.
%\label{average}
%\end{equation}

To implement this approach, we need to determine the equations
of motion for $\Theta_j(t), \Phi_j(t)$ which would produce a good approximation
for $p(\{\theta_j,\phi_j\},t)$. 
As a first step, let us find the exact
equations of motion for $p(\{\theta_j,\phi_j\},t)$. For simplicity, let us
study one term in the central-spin Hamiltonian (\ref{hamt}), i.e.\
we consider two spins 1/2 (the central spin and the $k$-th bath spin)
coupled by the isotropic Heisenberg interaction 
${\cal H}_k= A_k{\mathbf S}_0{\mathbf S}_k$. The most general form
for the two-spin density matrix is 
$\rho= w_{00}{\mathbf 1}_0{\mathbf 1}_k + w_{0\alpha}{\mathbf 1}_0\sigma^\alpha_k
+ w_{\beta 0}\sigma^\beta_0{\mathbf 1}_k+ w_{\lambda\nu}\sigma^\lambda_0\sigma^\nu_k$,
where $\alpha=x,y,z$ (and similarly for other Greek indices), and
$\sigma^\alpha_0$ and $\sigma^\beta_k$ denote the Pauli matrices for the $0$-th
and the $k$-th spin, respectively. Here and below, we assume summation over
the repeating indices. From the von Neumann's equation
$\dot\rho(t)=i[\rho(t),{\cal H}]$, we obtain
%the trivial equation 
$\dot w_{00}(t)=0$ (which expresses that ${\mathrm{Tr}}\rho(t)=1$),
and
\begin{eqnarray}
\nonumber
\dot w_{0\gamma}(t)&=& \frac{A_k}{2}\epsilon_{\alpha\beta\gamma} w_{\alpha\beta}(t),
\quad
\dot w_{\gamma 0}(t)= -\frac{A_k}{2}\epsilon_{\alpha\beta\gamma} w_{\alpha\beta}(t),\\
\dot w_{\alpha\beta}(t)&=& \frac{A_k}{2}\epsilon_{\alpha\beta\gamma} 
  [w_{\gamma 0}(t)-w_{0\gamma}(t)],
\label{exactrho}
\end{eqnarray}
where $\epsilon_{\alpha\beta\gamma}$ is a completely antisymmetric unity
tensor (permutation symbol). These equations of motion determine
the dynamics of $p(\{\theta_0,\phi_0,\theta_k,\phi_k\},t)$.
From the P-representation (\ref{prep}) it follows that  
$p(\{\theta_0,\phi_0,\theta_k,\phi_k\},t)=p_{00}(t) + p_{0\alpha}(t)c^\alpha_k
+ p_{\beta 0}(t) c^\beta_0 + p_{\lambda\nu}(t) c^\lambda_0 c^\nu_k$,
where $p_{00}=(1/4\pi^2)w_{00}$, $p_{0\alpha}=(3/4\pi^2)w_{0\alpha}$, 
$p_{\alpha 0}=(3/4\pi^2)w_{\alpha 0}$, and 
$p_{\alpha\beta}=(9/4\pi^2)w_{\alpha\beta}$.
Here, we used the shorthand notations $c^x_0=\sin{\theta_0}\cos{\phi_0}$,
$c^y_0=\sin{\theta_0}\sin{\phi_0}$, $c^z_0=\cos{\theta_0}$ (and similarly
for $c^x_k$, $c^y_k$, $c^z_k$).
The spherical harmonics of the order higher
than one can also be added to $p(\{\theta_0,\phi_0,\theta_k,\phi_k\})$, 
but they do not change the density matrix $\rho$,
and therefore are not physically significant.

The mean-field structure of the P-representation
for the density matrix (Eq.~\ref{prep}) suggests that the equations of
motion for $\{\Theta_0(t),\Phi_0(t),\Theta_k(t),\Phi_k(t)\}$ should also have a 
mean-field form corresponding to Eq.~\ref{eoms}, but the
local fields should be re-defined to provide optimal approximation
for ${\mathbf s}_0(t)$. For simplicity, we omit the discussion of the general
form for ${\mathbf h}_0(t)$ and ${\mathbf h}_k(t)$, and proceed to the
answer. We introduce
the shorthand notations $C^x_0=\sin{\Theta_0}\cos{\Phi_0}$,
$C^y_0=\sin{\Theta_0}\sin{\Phi_0}$, $C^z_0=\cos{\Theta_0}$ (and similarly
for $C^x_k$, $C^y_k$, $C^z_k$), and postulate the following equations of motion:
\begin{equation}
\dot {\mathbf C}_0 = g_1[{\mathbf C}_k\times {\mathbf C}_0],\quad
\dot {\mathbf C}_k = -g_2[{\mathbf C}_k\times {\mathbf C}_0]
\label{eoms1}
\end{equation}
(cf.\ Eq.~\ref{eoms}), where $g_1$ and $g_2$ are to be determined.
By substituting these equations into the probability
distribution 
$p(\{\Theta_0(t),\Phi_0(t),\Theta_k(t),\Phi_k(t)\})=P_{00} + P_{0\alpha}C^\alpha_k
+ P_{\beta 0}C^\beta_0 + P_{\lambda\nu}C^\lambda_0 C^\nu_k$, and
using the P-representation (\ref{prep}), we
obtain the following equations of motion for the density matrix $\rho$:
\begin{eqnarray}
\nonumber
\dot w_{0\gamma}(t)&=& -g_2 \epsilon_{\alpha\beta\gamma} w_{\alpha\beta}(t),
\quad
\dot w_{\gamma 0}(t)= g_1 \epsilon_{\alpha\beta\gamma} w_{\alpha\beta}(t),\\
\dot w_{\alpha\beta}(t)&=& (1/3)\epsilon_{\alpha\beta\gamma} 
  [g_2 w_{0\gamma}(t)- g_1 w_{\gamma 0}(t)],
\label{tdmfrho}
\end{eqnarray}
(cf.\ Eqs.~\ref{exactrho}), and, trivially $\dot w_{00}(t)=0$.
The Eqs.~\ref{exactrho} and \ref{tdmfrho} are incompatible for any 
$g_1$ and $g_2$ (i.e., TDMF is never exact). However, we are interested
only in $w_{\gamma 0}(t)$, since only this term determines the value
of ${\mathbf s}_0(t)$. Therefore, we choose 
$g_1=g_2=g=A_k \sqrt{3}/2$, and differentiate Eqs.~\ref{exactrho} and \ref{tdmfrho} 
with respect to time once more. Then we see that both Eqs.~\ref{exactrho} 
and \ref{tdmfrho} produce the same result:
%\begin{equation}
$\ddot w_{\gamma 0}(t) = -\ddot w_{0\gamma}(t) = (A_k^2/2)(w_{0\gamma}-w_{\gamma 0})$,
%\end{equation}
so that the functions $w_{\gamma 0}(t)$ produced by the approximate equations
(\ref{eoms1}) and by the exact equations (\ref{exactrho}) coincide,
provided that the initial conditions $w_{\gamma 0}(0)$ and 
$\dot w_{\gamma 0}(0)$ also coincide. The latter condition
is satisfied when all $w_{\alpha\beta}(0)=0$, so 
the method described above is applicable only to unpolarized
baths.
On the other hand, for polarized baths, one can use the perturbational
approaches \cite{loss,hu}, so this limitation is not serious. 
%Therefore, in order 
%to fix the standard TDMF in case of spins 1/2, we just need to take Eqs.~\ref{heff}, 
%and replace $A_k$ by $A_k\sqrt{3}/2$ \cite{notesemi}.
%
%
\begin{figure}
\includegraphics[width=4cm]{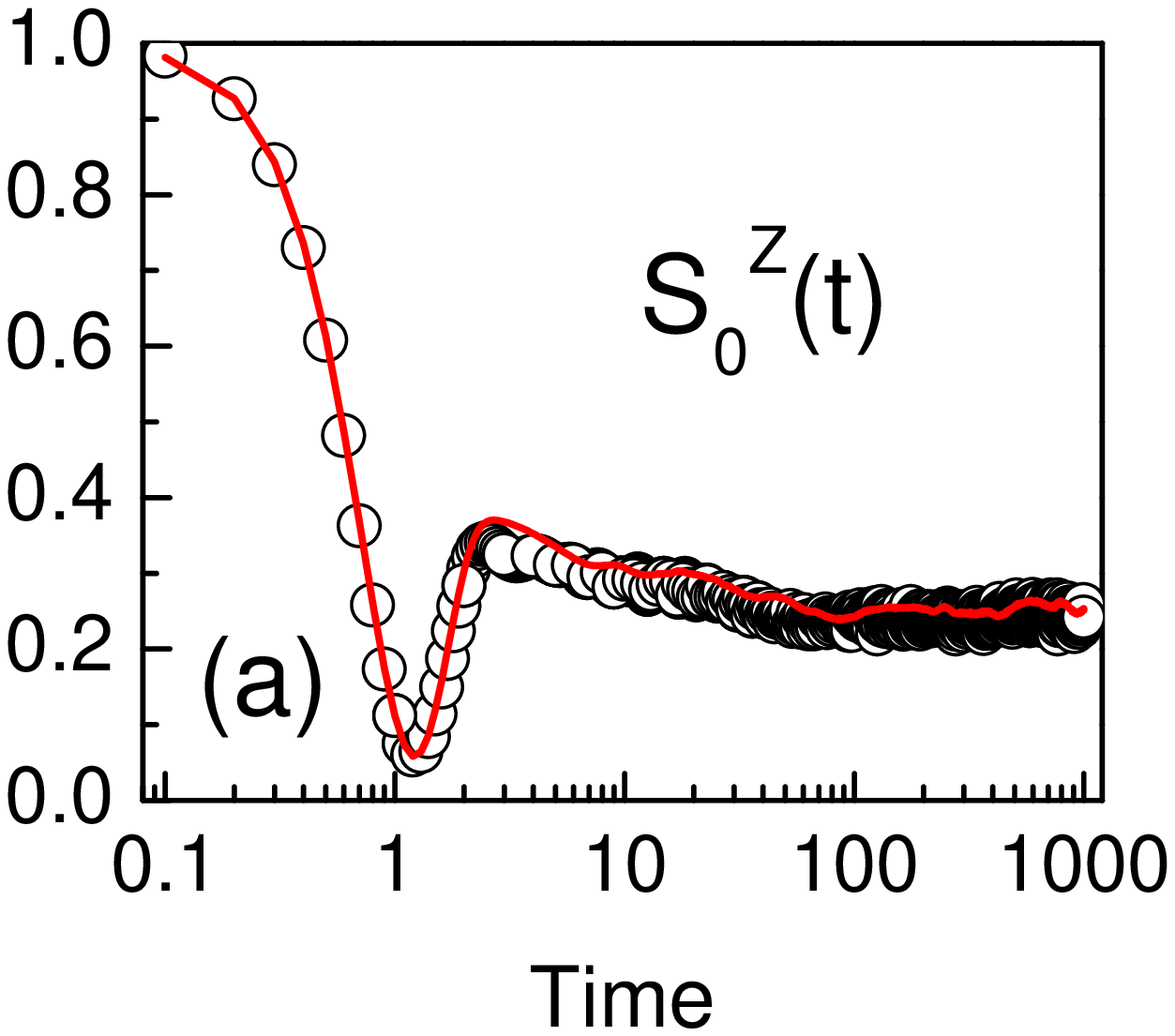}
\includegraphics[width=4cm]{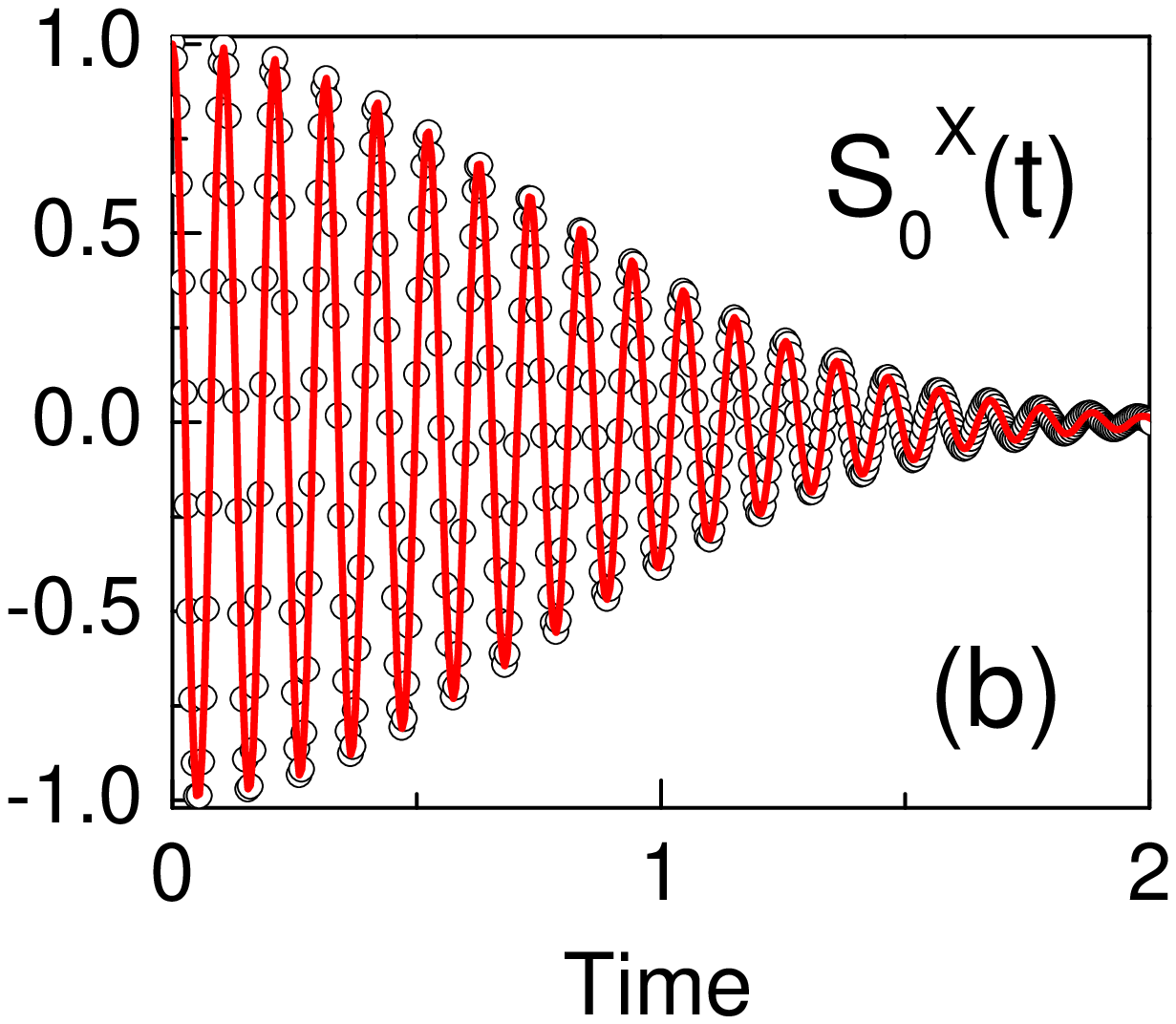}
\caption{\label{fig1} Longitudinal $s_0^z(t)$ (a) and 
transversal $s_0^x(t)$ (b) relaxation of the central
spin 1/2 coupled to a bath of $N=21$ spins 1/2. Couplings $A_k$ are
randomly distributed between 0 and 1.0, external fields are $H_0=0$ (a),
and $H_0=60.0$ (b). Solid lines --- exact solutions, open circles --- approximation. 
Agreement is excellent.}
\end{figure}
Therefore, in order to fix the standard TDMF in case of spins 1/2, we 
just need to take Eqs.~\ref{heff}, and replace $A_k$ by $A_k\sqrt{3}/2$.
It may seem that we just derived the standard semiclassical equations
of motion, but this is not correct. For instance, if we take 
unequal spins 1 and 1/2 ($S_0=1$ and $S_k=1/2$) then the analysis above
gives $g_1=A_k\sqrt{3}/2$, $g_2=A_k\sqrt{7}/2$, while the semiclassics
would give $g_1=A_k\sqrt{3}/2$, $g_2=A_k\sqrt{2}$. Moreover, the initial
conditions in our approach and in the semiclassical approach 
are different. For example, in case when the central
spin 1/2 is initially directed along the $z$-axis, so that the initial density
matrix 
$\rho(0)=2^{-N} |\uparrow\rangle\langle\uparrow|\bigotimes_{k=1}^N {\mathbf 1}$,
our approach requires 
$p(\{\theta_j,\phi_j\},0)=(4\pi)^{-N-1}(1+3\cos{\theta_0})$, while
the semiclassical approximation would correspond to
$p(\{\theta_j,\phi_j\},0)=(4\pi)^{-N-1}\delta(\cos{\theta_0}-1)$,
where $\delta(\dots)$ is the Dirac's delta-function.
\begin{figure}
\includegraphics[width=4cm]{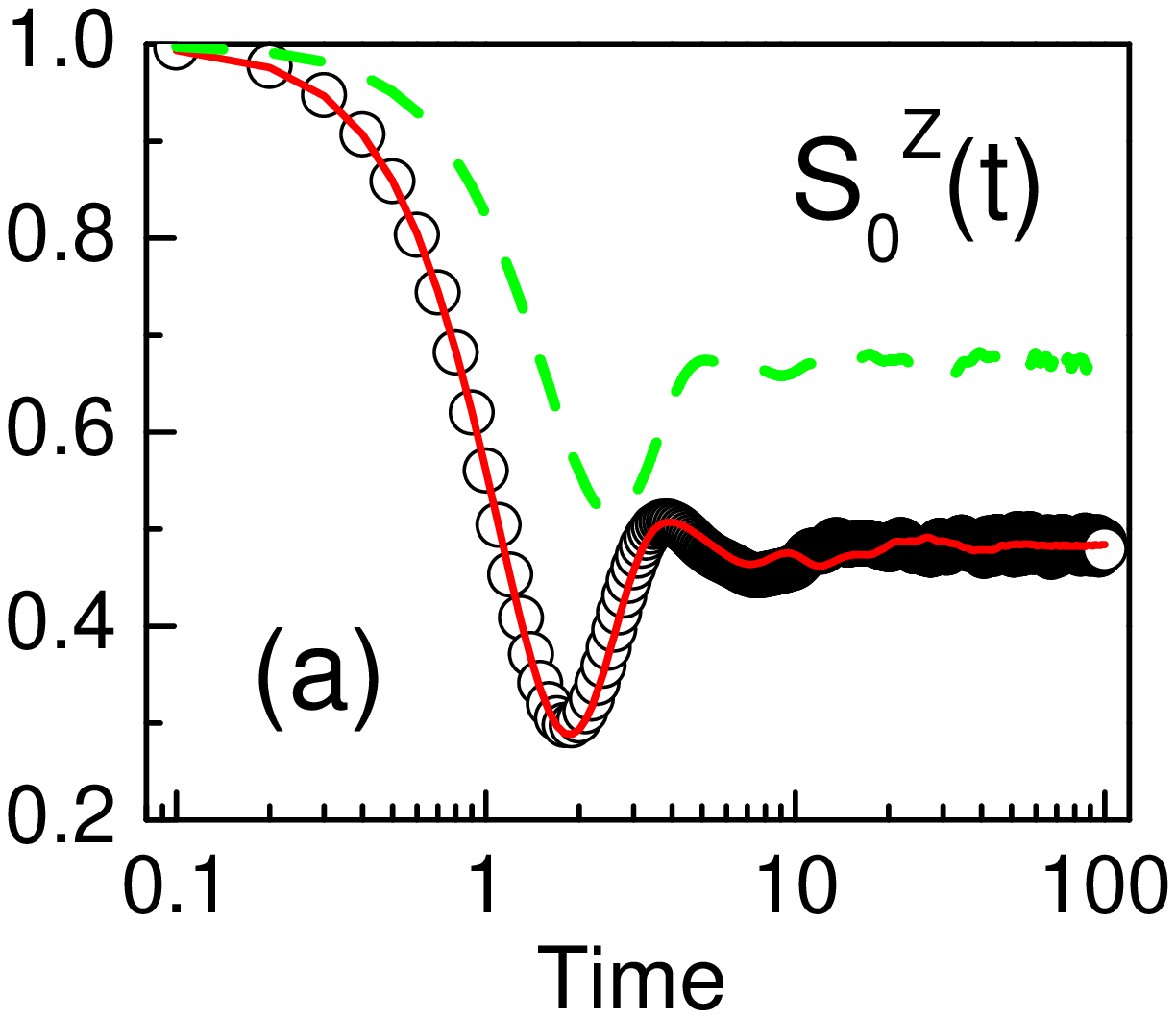}
\includegraphics[width=4cm]{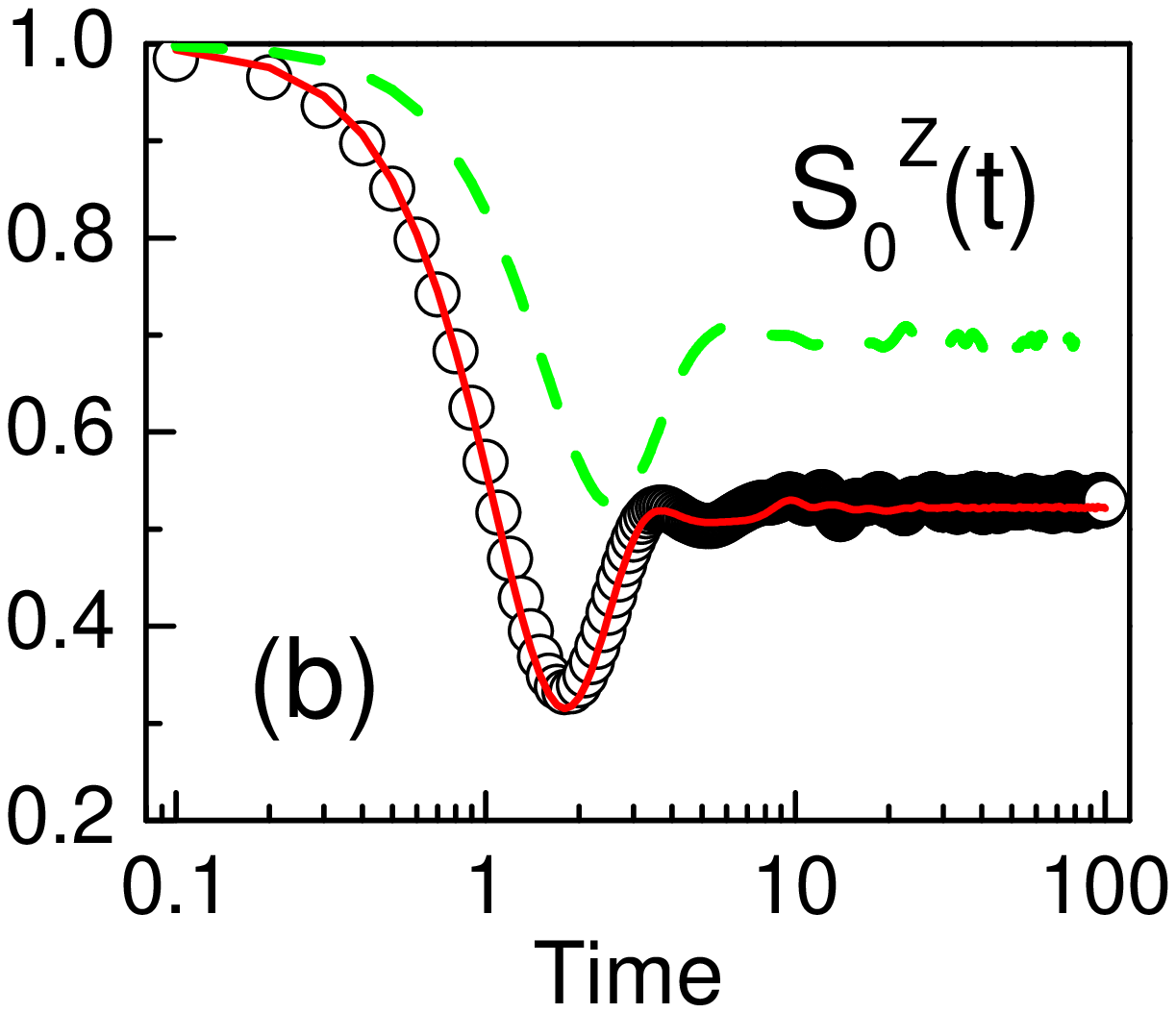}
\caption{\label{fig2} Longitudinal relaxation $s_0^z(t)$ of the central
spin, couplings $A_k$ are
randomly distributed between -0.4 and 0.6, field $H_0=1.0$.
(a) central spin 1/2, $N=21$ bath spins (b) central spin 1, $N=19$ bath spins. 
Solid lines --- exact solutions, open circles --- our approximation;
agreement is excellent. Dashed lines --- standard TDMF; the disagreement
with the exact solution is significant.}
\end{figure}

The approach developed above gives excellent results at both
short and long times, for
different distributions of the coupling constants $A_k$, 
external fields $H_0$, and is applicable to both longitudinal
and transversal relaxation. A small fraction of the representative test results
for a moderate number of bath spins ($N=15$--20), 
is shown in Figs.~\ref{fig1},\ref{fig2}, where we used the
Chebyshev expansion method \cite{ddr} for exact solution of the quantum 
problems. 
The longitudinal decay shown in Fig.~\ref{fig1} is typical;
the long-time tail suggests the slow relaxation $s_0^z(t)\sim 1/\ln{t}$,
but the results are not conclusive, since for moderate-sized systems
$s_0^z(t\to\infty)$ saturates at a value far from zero. Thus, we
have to study larger systems and to use the approximate method.
\begin{figure}
\includegraphics[width=4cm]{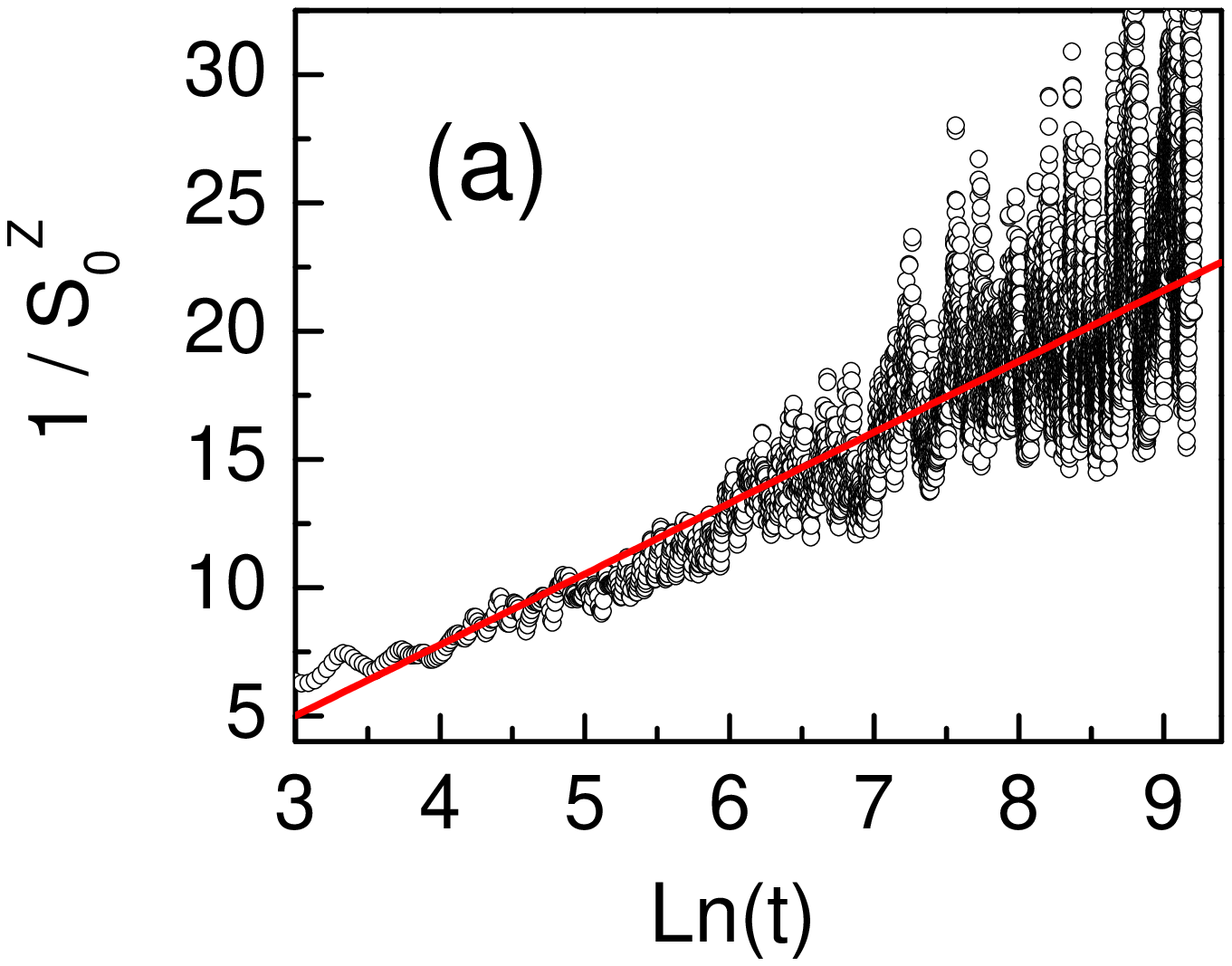}
\includegraphics[width=4cm]{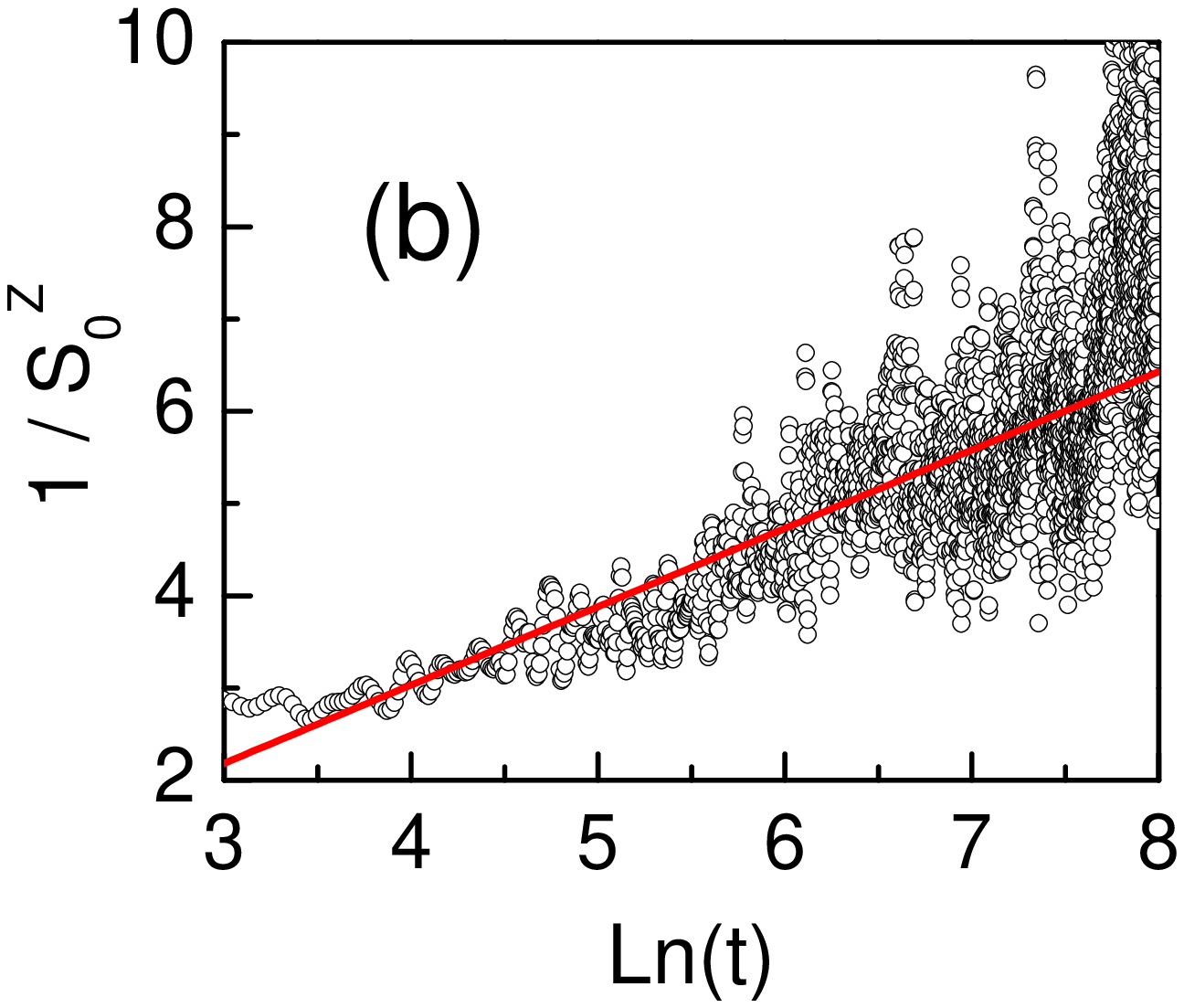}
\caption{\label{fig3} Long-time relaxation of the central
spin 1/2 coupled to a bath of 16000 spins 1/2, field $H_0=0$.
Graphs show $1/s_0^z(t)$ as a function of $\ln{t}$.
The coupling constants were calculated as
$A_k= (1/14) u({\mathbf x}_k)$, where the $u({\mathbf x})$ is the electron density.
(a): $u({\mathbf x})$ is taken as a Gaussian
with the half-widths $d_x=8.4a$, $d_y=9.1a$, $d_z=2.2a$ ($a$ is the
lattice parameter), shifted from the center of the lattice by the vector
$(0.252a,0.448a,0.1a)$;
(b): $u({\mathbf x})$ is taken as an exponential function of ${\mathbf x}$, 
with the same parameters. We used an extra averaging
over 20 neighboring time points to decrease the number of realizations. 
The solid lines are obtained from raw data.}
\end{figure}

\begin{figure}
\includegraphics[width=4cm]{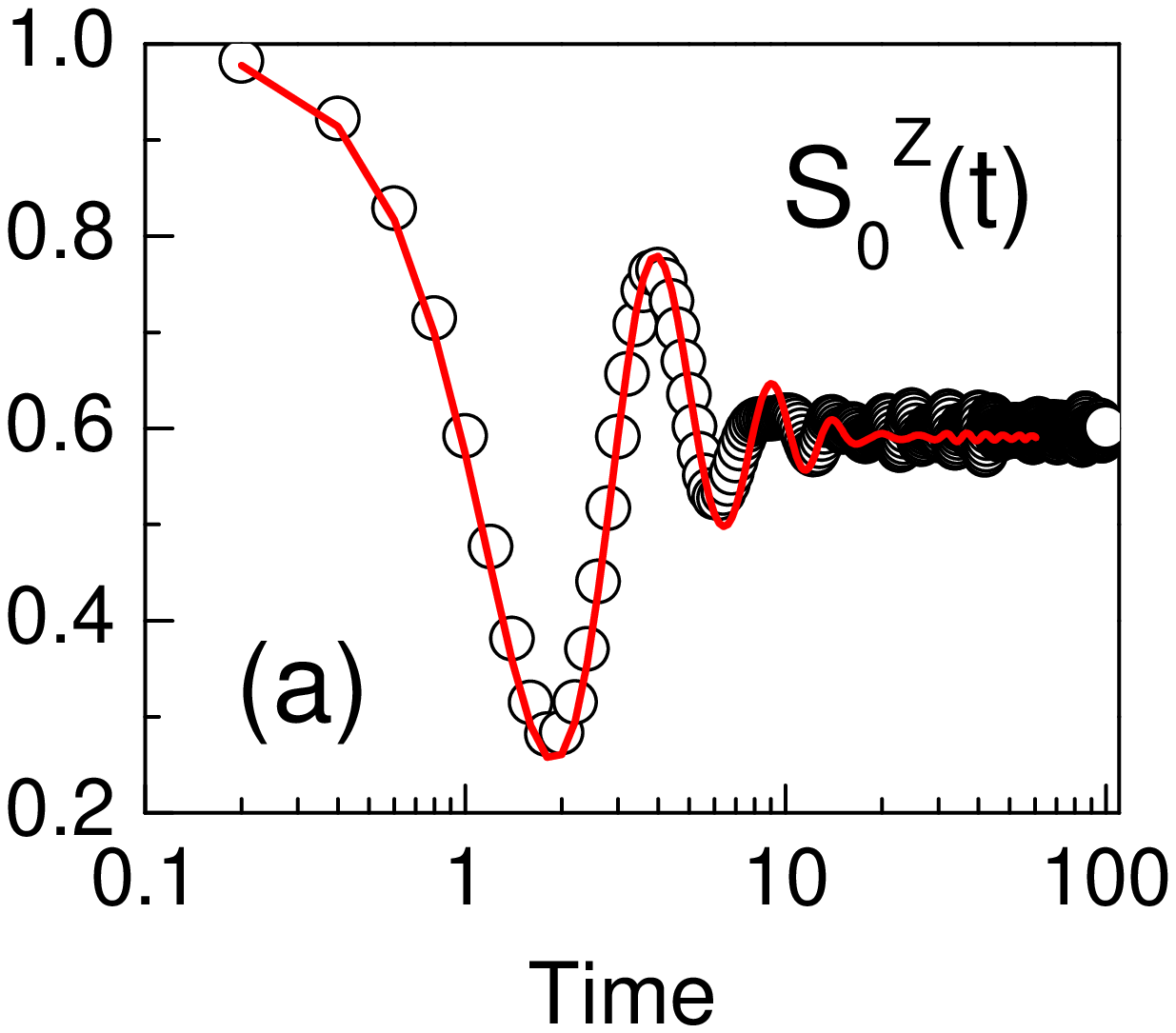}
\includegraphics[width=4cm]{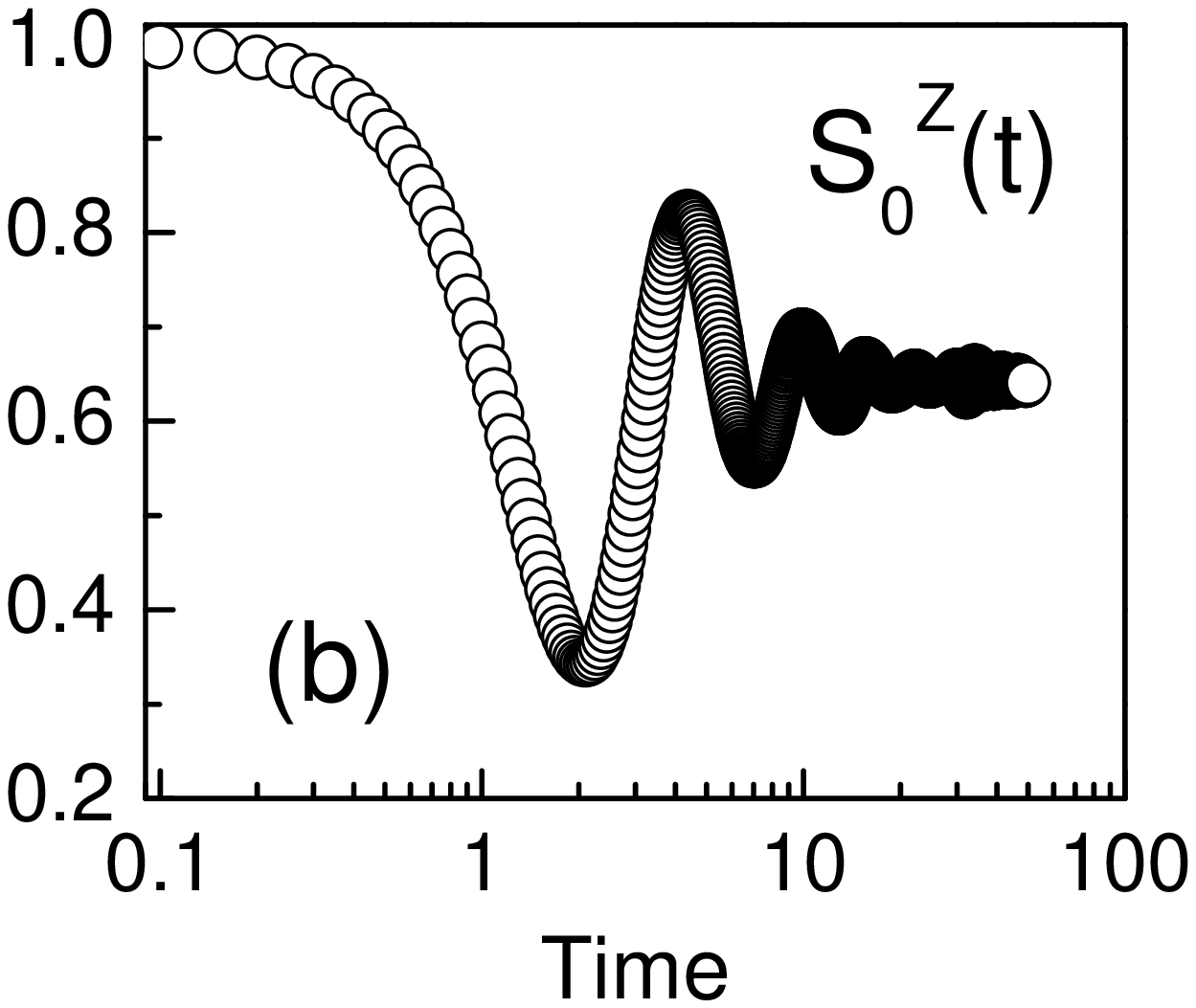}
\caption{\label{fig4} Longitudinal relaxation $s_0^z(t)$ of the central
spin 1/2 coupled to a bath of spins 1/2 via the anisotropic $X$-$Y$
Hamiltonian, external field $H_0=1.0$. (a): $N=21$ bath spin,
$A_k$ are randomly distributed between 0 and 1.0. Solid 
line --- exact solution, open circles --- approximation. (b): $N=2000$ 
spins, the couplings $A_k$ are the same as in Fig.~\ref{fig3}a,
but for smaller lattice $N_x=N_y=20$, $N_z=5$.}
\end{figure}

We considered the systems with $N=1000$-15000 spins.
For small $N=2$--5, our approximation
is very crude because the equations
of motion (\ref{eoms1}) lead to the appearance of higher-order 
spherical harmonics
in the function $p(\{\theta_j,\phi_j\},t)$ (the terms 
proportional to $c_0^\alpha c_0^\beta c_k^\lambda c_k^\nu$ etc.).
These terms do not change the form of $\rho(t)$ (they disappear
after integration in Eq.~\ref{prep}), but they affect the equations
of motion for physically relevant terms, i.e.\ the actual equation
of motion for $w_{\gamma 0}(t)$ becomes 
$\ddot w_{\gamma 0}(t) = (A_k^2/2)(w_{0\gamma}-w_{\gamma 0}) + V$,
where $V$ is the contribution from the higher-order harmonics. However,
the contribution of $V$ into ${\mathbf s}_0(t)$ is bounded, and 
quickly decreases for larger $N$, so we expect our approximation 
for ${\mathbf s}_0(t)$ to work the better the larger $N$ is. This is a natural
expectation, since our method is based on TDMF, which works better
for larger number of bath spins coupled to the central spin. Moreover,
this assumption is well-confirmed by numerous numerical tests.
%
%To avoid negative probabilities, we used the initial
%distribution $p(\{\theta_j,\phi_j\},0)=(4\pi)^{-N-1}(1+\cos{\theta_0})$,
%corresponding to 
%$\rho(0)=2^{-N-1}({\mathbf 1}+(2/3)|\uparrow\rangle\langle\uparrow|)
%\bigotimes_{k=1}^N {\mathbf 1}$,
%and calculated ${\mathbf s}_0(t)=3{\mathrm{Tr}}\rho(t){\mathbf S}_0$;
%this transformation is exact, since the identity density matrix
%does not change under Hamiltonian evolution. 
%
In Fig.~\ref{fig3},
we present the long-time longitudinal relaxation of an electron spin
in a model QD; we assumed that the bath spins 1/2 are placed
at the sites of a piece of a cubic lattice with the size $N_x\times N_y\times N_z$, 
($N_x=N_y=40$, $N_z=10$, so the total number of bath spins
$N=N_x N_y N_z=16000$). The long-time relaxation is extremely slow,
clearly demonstrating the law $1/\ln{t}$ (see \onlinecite{nazarovComment}). 
Our simulations show that the law
$1/\ln{t}$ holds for unpolarized baths for different forms of the
electron densities, i.e.\ for different distributions of $A_k$
(two examples are given in Fig.~\ref{fig3}). 
Our approach also performs well (see Fig.~\ref{fig4})
for an anisotropic $X$-$Y$
coupling between the central and the bath spins
${\cal H}=H_0S^z_0 + \sum_k A_k (S^x_0 S^x_k + S^y_0 S^y_k)$,
which is important for analyzing interesting experiments
of Ref.~\onlinecite{taylor}.
The results show good qualitative
agreement with the experimental curves \cite{taylor}, but the experiments
are performed with $\sim 5$--10\% polarization, so that our model needs
further development for rigorous quantitative comparison.

Summarizing, we used the spin coherent states P-representation to
develop a novel approach to the quantum central 
spin problem, based on the time-dependent mean field theory.
The approach gives excellent agreement with the exact solutions, and
is valid for a wide range of systems and conditions.
We use it to study the long-time longitudinal relaxation
of the electron spin in a quantum dot, and find that the slow decay
$1/\ln{t}$ is observed in different situations. Our approach 
provides an interesting extension of the mean-field theory,
and is applicable to many-spin central systems as well.

The authors would like to thank L.~Glazman, M.~Lukin, A.~Polkovnikov, and 
J.~Taylor for helpful suggestions and discussions. This work was supported 
by the NSA and ARDA under Army Research
Office (ARO) contract DAAD 19-03-1-0132.  K.A. and E.D. were supported by the NSF 
grant DMR-0454504.


\begin{thebibliography}{99}
\bibitem{johnson} A. C. Johnson {\it et al.}, 
  Nature {\bf 435}, 925 (2005); F. H. L. Koppens {\it et al.}, Science {\bf 309}, 1346 (2005); J. R. Petta {\it et al.}, 
  Science {\bf 309}, 2180 (2005).
\bibitem{taylor} J. M. Taylor (private communication).
\bibitem{gaudin} M. Gaudin, J. de Physique {\bf 37}, 1087 (1976).
\bibitem{khaetskii} A. V. Khaetskii, D. Loss, and L. Glazman, Phys. Rev. Lett.
  {\bf 88} 186802 (2002).
\bibitem{efros} I. A. Merkulov, Al. L. Efros, and M. Rosen, Phys. Rev. B {\bf 65},
  205309 (2002).
\bibitem{loss} W. A. Coish and D. Loss, Phys. Rev. B {\bf 70}, 195340 (2004).
\bibitem{hu} Changxue Deng and Xuedong Hu, cond-mat/0510379
%
%\bibitem{notesemi}It may seem that we just derived the standard semiclassical equations
%of motion, but this is not correct. For instance, if we take 
%unequal spins 1 and 1/2 ($S_0=1$ and $S_k=1/2$) then the analysis above
%gives $g_1=A_k\sqrt{3}/2$, $g_2=A_k\sqrt{7}/2$, while the semiclassics
%would give $g_1=A_k\sqrt{3}/2$, $g_2=A_k\sqrt{2}$. Moreover, the initial
%conditions in our approach and in the semiclassical approach 
%are different. For example, in case when the central
%spin 1/2 is initially directed along the $z$-axis, so that the initial density
%matrix 
%$\rho(0)=2^{-N} |\uparrow\rangle\langle\uparrow|\bigotimes_{k=1}^N {\mathbf 1}$,
%our approach requires 
%$p(\{\theta_j,\phi_j\},0)=(4\pi)^{-N-1}(1+3\cos{\theta_0})$, while
%the semiclassical approximation would correspond to
%$p(\{\theta_j,\phi_j\},0)=(4\pi)^{-N-1}\delta(\cos{\theta_0}-1)$,
%where $\delta(\dots)$ is the Dirac's delta-function.
%
\bibitem{dkdh} V. V. Dobrovitski {\it et al.}, Phys. Rev. Lett. {\bf 90}, 210401 (2003).
\bibitem{akakii} A. Melikidze {\it et al.}, Phys. Rev. B {\bf 70}, 014435 (2004)
\bibitem{nazarov} S. I. Erlingsson and Y. V. Nazarov, Phys. Rev. B {\bf 70},
  205327 (2004). 
\bibitem{emil} E. A. Yuzbashyan {\it et al.}, Phys. Rev. B {\bf 72}, 144524 (2005);  
  E. A. Yuzbashyan {\it et al.}, J. Phys. A {\bf 38}, 
  7831 (2005). Note that these works deal with classical systems.
\bibitem{dirac} P. A. M. Dirac, Proc. Cambridge Phil. Soc. {\bf 26}, 376 (1930). 
\bibitem {gardiner} R. J. Glauber, Phys. Rev. Lett. {\bf 10}, 84 (1963);
  C. W. Gardiner and P. Zoller, {\it Quantum Noise}
  (Springer-Verlag, Berlin, Heidelberg, New York, 2000).
\bibitem{ddr} V. V. Dobrovitski and H. A. De Raedt, Phys. Rev. E {\bf 67}, 
  056702 (2003).
  %; H. De Raedt and V. V. Dobrovitski, in {\it Computer
  %Simulation Studies in Condensed-Matter Physics XVI\/}, D. P. Landau,
  %S. P. Lewis, and H.-B. Sch\"uttler (eds.) (Springer Verlag, Berlin,
  %2004).
\bibitem{klauder} J. R. Klauder, B.-S. Skagerstam {\it Coherent States\/}
  (World Scientific, Singapore, 1985).
\bibitem{nazarovComment} Similar relaxation was found in \onlinecite{nazarov}
for 2-D Gaussian electron density, using adiabatic approximation applied 
to the semiclassical equations of motion. Accuracy of this method
is unclear; it is supposed to be correct for large $N$, but
we can neither confirm nor reject that: the
exact solution for more than $\sim 20$ bath spins is extremely difficult, 
while for $N=22$ bath spins the exact solution disagrees with 
the approximation of Ref.~\onlinecite{nazarov}. Also, $1/\ln{t}$ was predicted 
from the perturbation theory for {\it polarized\/} baths \cite{loss} in case of 
2-D dot with Gaussian electron density, but other forms of decay
have been predicted for other electron density distributions.
In our simulations for {\it unpolarized\/} baths, we see $1/\ln{t}$
for different electron density distributions. 
\end{thebibliography}
\end{document}